\documentclass{aa}  
\usepackage{graphicx}
\usepackage{txfonts}
\usepackage[usenames, dvipsnames]{color}
\begin{document} 

\title{Globular clusters in the Galactic center region: Expected behavior within the infall and merger scenario}

   %\subtitle{I. Overviewing the $\kappa$-mechanism}

\author{Maria Gabriela Navarro \inst{1} \and
Roberto Capuzzo-Dolcetta  \inst{2} \and
Manuel Arca-Sedda  \inst{3} \and
Dante Minniti \inst{4,5} }
 \institute{ INAF—Osservatorio Astronomico di Roma, Via di Frascati 33, 00040, Monteporzio Catone, Italy \\
\email{maria.navarro@inaf.it}
 \and
Dipartimento di Fisica, Universit\`a degli Studi di Roma ``La Sapienza'', P.le Aldo Moro, 2, I00185 Rome, Italy 
  \and
 Department of Physics and Astronomy, University of Padova, Italy
   \and
 Departamento de Ciencias F\'isicas, Facultad de Ciencias Exactas, Universidad Andr\'es Bello, Av. Fern\'andez Concha 700, Las Condes, Santiago, Chile
 \and
Vatican Observatory, V00120 Vatican City State, Italy
}

% \abstract{}{}{}{}{} 
% 5 {} token are mandatory
 
\date{Received xxx; Accepted xxx}
\abstract{
    In this work, we reexamine the infall and merger scenario of massive clusters in the Milky Way's potential well as a plausible Milky Way formation mechanism. We aim to understand how the stars of the merging clusters are redistributed during and after the merger process. We used, for the first time, high-resolution  simulations with concentrated in the 300 pc around the Galactic center. 
      We adopted simulations developed in the framework of the Modelling the Evolution of Galactic Nuclei (MEGaN) project.
    We compared the evolution of representative clusters in the mass and concentration basis in the vicinity of a supermassive black hole. 
    We used the spatial distribution, density profile, and the $50\%$ Lagrange radius (half mass radius) as indicators along the complete simulation to study the evolutionary shape in physical and velocity space and the final fate of these representative clusters. 
    We find that the least massive clusters are quickly (<10 Myr) destroyed. 
    On the other hand, the most massive clusters have a long evolution, showing variations in the morphology, especially after each passage close to the supermassive black hole. The deformation of the clusters depends on the concentration, with general deformations for the least concentrated clusters and outer strains for the more concentrated ones. 
    At the end of the simulation, a dense concentration of stars belonging to the clusters was formed.
    The particles that belong to the most massive and most concentrated clusters are concentrated in the innermost regions, meaning that the most massive and concentrated clusters contribute a more significant fraction of particles to the final concentration. This finding suggests that the population of stars of the nuclear star cluster formed through this mechanism comes from massive clusters rather than low-mass globular clusters. \\
} 

\maketitle
%
%-------------------------------------------------------------------

\section{Introduction}
    Several mechanisms have been proposed within the framework of galactic formation.
    Among the most widely accepted scenarios,  the monolithic collapse model is notable, given its support of galactic formation via the collapse of a massive gas cloud. This scenario was first proposed by \cite{ELS} and it now known as the ELS model. 
Another proposed scenario is the hierarchical model, a generic feature of cold dark matter (CDM) models, which suggests that the galaxies were formed by building blocks that formed first and then, the Galaxy was formed from smaller galaxies carrying globular clusters (GCs) with them, and deposited them mainly in the outskirts of the Galactic halo \citep{wr78,Baugh96,Neistein06}. 
     Overall, GCs are the oldest stellar systems known and they provide information  that is vital to improving our understanding the earliest stages of the Milky Way.
   
    From an observational point of view, the census of GCs in the Milky Way increased in number in the last years, reaching 156 GCs \citep{harris10}, with many more candidates to be confirmed \citep{dante17,elisa22a,elisa22b}.
    One of the most critical limitations of studying GCs is extinction, which prevents the detection of clusters in the densest areas, such as the innermost region of the Milky Way, where we expect to find them and also their remnants because the potential well of the Galaxy is deeper and the disruptive dynamical effects should be stronger. 
     This situation is improved with the new era of infrared surveys such as 2MASS \citep{Skrutskie06} and VVV Survey \citep{vvvminniti10}. 
    \cite{Minniti21} investigated the GCs in the innermost regions of the Milky Way. They confirmed that VVV-GC002 is the closest known GC to the Galactic center. This metal-rich GC located at only 1.1 deg from the Galactic center, equivalent to $R_G = 0.4$ kpc, was discovered initially by \cite{Moni11}.
 
    \cite{Minniti21} also found that there appears to be a forbidden zone of radius $R_G \sim 0.1$ kpc around the Galactic center, where GCs are crushed, and only young clusters can be seen. These young clusters would presumably not last long \citep{Habibi13, Habibi14, Hosek15, Rui19,Libralato20,Libralato22}. 
        The debris of destroyed GCs might be found in this zone of tidal disruption surrounding the supermassive black hole (BH). Some of these destroyed primordial GCs (hereafter PGCs) may have helped build the massive nuclear star cluster at the center of our Galaxy \citep{rcd93, Arca14}. 
This scenario is also supported by the presence of RR Lyrae stars in this region \citep{d95, navarro21}, which are excellent indicators for  studying the spatial distribution and dynamics of the population that comes from these GCs.% that acted as building blocks of the Galaxy's central region.

    On the theoretical side, there have been numerous GC simulations that consider the different dynamical processes that affect their evolution \citep{Chandrasekhar42, Henon61, Fall77,Fall85, Larson70, tre75, Tremaine84,Aarseth74, Aarseth99, Heggie79, Heggie14, Chernoff90, Vesperini97,Gnedin97, Gnedin14, Baumgardt02, Baumgardt03, Gieles06, Carlberg17, Carlberg18, Khoperskov18}.
    In particular, we are interested in the evolutionary shape in physical and velocity space and their survival and destruction mechanisms in the presence of a single supermassive BH, such as the one located at the center of the Milky Way. This study is crucial for its subsequent comparison with the population of clusters around the Galactic center, which has been increasing in recent years thanks to the new generation of infrared telescopes, which will further increase thanks to new missions such as the James Webb Space Telescope (JWST) and the Roman Space Telescope. (WFIRST, \citealt{Green12, Spergel15})

    This paper presents new simulations for GCs in the Galactic center region. In Section 2, the details of the model and the initial conditions used are presented. Section 3 presents the method to study the evolution and morphology of the representative GCs in the vicinity of the supermassive BH. Our results are discussed in Section 4, and in Section 5, our findings are compared with the observations. Finally, our conclusions are presented in Section 6. 
    
    \section{Galactic model and initial conditions}
We used the HiGPUs code \citep{rcd13}, an N-body code suitable for studying the dynamical evolution of stellar systems composed of up to 10 million stars with precision guaranteed by direct summation of the pair-wise forces.
    The code is written by combining C and C++ programming languages, and it is parallelized using a message passing interface (MPI), along with OpenMP and OpenCL to allow for the utilization of GPUs of different vendors.
    The code implements the Hermite's sixth order time integration scheme \citep{Nitadori08} with block time steps, allowing for high levels of precision and speed in studying the dynamical evolution of star systems.
    The coarse-grained parallelization establishes a one-to-one correspondence between MPI process and computational nodes, and each MPI process manages all the GPUs available per node. 
    
    For this analysis, we used simulations of GCs decaying to the center of the Galaxy (\cite{manuel17}, \cite{manuel18}).
    The simulation is performed in the framework of the Modelling the Evolution of Galactic Nuclei (MEGaN) project and consists of studying the evolution and merger of 41 GCs in the presence of a supermassive BH.  We used a Dehnen model \citep{Dehnen93} to model the Galactic bulge. The number of GCs is an arbitrary choice based on the limiting computing facilities.
       
    The total number of particles is $N = 2^{20} = 1,048,576$, which includes the Supermassive BH of $M_{SMBH}= 4.5 \times 10^6  M_\odot$, $N= 478,107$ particles belonging to the 41 GCs with stellar masses of $M= 92  M_\odot$ and $N= 570,468$ background particles with particle masses of $M= 180  M_\odot$ each. Due to computational limitations, the field particles are twice as massive as the cluster particles. 
    The GCs have different density distributions and cover a range from $N = 1477$ to $N = 21383$ particles, that is, from $M_{GC} = 10^5 M_\odot$ to $M_{GC} = 2 \times 10^6 M_\odot$ and core radii from $R_c = 0.2$ pc to $R_c = 1.7$ pc, which is the radius where the density has dropped to half the central value.
       Table~\ref{tab:M_tablec} presents the main characteristics (core radius, number of particles, and total mass) of the GCs used in this simulation. 
 The system's total mass is $M= 1.5 \times 10^8 M_\odot$ and the whole sample is initially confined within a radius of 300 pc, which is justified by previous effects of dynamical friction.

\begin{table}
\caption{    Properties of the Globular clusters in the simulation, Column 1: GC name. Column 2: Number of particles. Column 3: Total mass in $10^6 \times M_\odot$. Column 4: Core radius in pc. }  
\label{tab:M_tablec}    
\centering                  
\begin{tabular}{c c  c c}      
\hline\hline          
     GC name & N & $M_{GC}$ & $R_c$ \\
     & & {\tiny ($10^6 M_\odot$)} & {\tiny (pc)} \\
\hline  
    1 & 18054 & 1.66 & 1.7\\
    2  & 12860 & 1.18 & 0.3\\
    3 & 1477 & 0.13 & 0.7\\
    4 & 20700 & 1.90 & 0.2\\
    5  & 17980 & 1.65 & 1.1\\
    6 & 10366 & 0.95 & 0.4\\
    7 & 5172 & 0.47 & 0.4\\
    8 & 18993 & 1.7 & 0.6\\
    9 & 9802 & 0.90 & 0.6\\
    10 & 2183 & 0.20 & 0.6\\
    11 & 17753 & 1.63 & 0.4\\
    12 &  4495 & 0.41 & 0.7\\
    13 & 18910 & 1.73 & 0.9\\
    14 & 20051 & 1.84 & 0.5\\
    15 &  17258 & 1.58 & 0.8\\
    16 & 3416 & 0.31 & 0.9\\
    17 &  4463 & 0.41 & 0.3\\
    18 &  5893 & 0.54 & 0.2\\
    19 & 13701 & 1.26 & 0.4\\
    20 & 14507 & 1.33 & 0.3\\
    21 & 14244 & 1.31 & 1.2\\
    22 & 18256 & 1.67 & 1.5\\
    23 & 21383 & 1.96 & 0.6\\
    24 & 7635 & 0.70 & 0.5\\
    25 & 11127 & 1.02 & 0.5\\
    26 & 11118 & 1.02 & 0.6\\
    27 &  15858 & 1.45 & 0.3\\
    28 &  5376 & 0.49 & 0.2\\
    29 &  10459 & 0.96 & 0.8\\
    30 &  1919 & 0.17 & 0.5\\
    31 &  11134 & 1.02 & 0.2\\
    32 &  6577 & 0.60 & 0.3\\
    33 &  3411 & 0.31 & 0.2\\
    34 &  2580 & 0.23 & 0.5\\
    35 &  17040 & 1.56 & 1.0\\
    36 &  8014 & 0.73 & 0.5\\
    37 &  14742 & 1.35 & 0.5\\
    38 &  19018 & 1.74 & 0.8\\
    39 &  17186 & 1.58 & 1.1\\
    40 &  8388 & 0.77 & 0.6\\
    41 &  14608 & 1.34 & 0.9\\
\hline 
\end{tabular}
\end{table}
   
    The simulation evolves during $\sim 200$ Myr with timestamps of $\Delta T = 0.083125$ Myr within a radius of 300 pc ($\sim$ 2.1 deg).
    No external potential is applied in this simulation because the nuclear bulge is modeled in a self-consistent way with field stars. 
    Further details of the simulations are presented in detail in our following article (Capuzzo-Dolcetta, Navarro, et al. in prep.).
    
\begin{figure}[b]
    \begin{center}
    \includegraphics[width=0.9\linewidth]{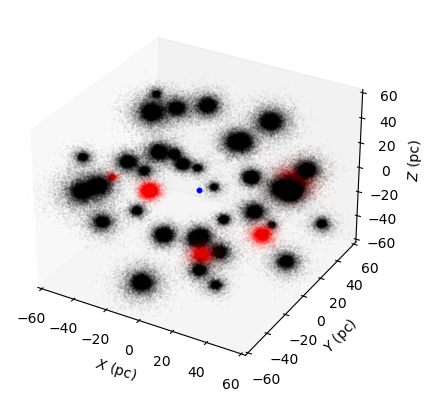}\\
    \caption
    {First snapshot of the simulation plotted on a 60 pc scale, including the 41 Globular clusters with a wide range of masses. 
    The blue dot corresponds to the supermassive BH in the center.
    The red clusters correspond to the clusters analyzed in this article selected according to mass and density, which act as representative samples for the analysis. 
    \label{fig:M_clblue}}
    \end{center}
    \end{figure}

    Our first and foremost assumption is that the supermassive BH was already in place when these PGCs merged. The outcome of the simulations is that all of these PGCs no longer exist, having formed the nuclear star cluster. Figure~\ref{fig:M_clblue} shows the 3D spatial distribution of the first snapshot of the simulation, where the 41 GCs begin to fall in the potential well of the Galaxy generated by background particles (which are not included in the figure). The blue dot represents the supermassive BH at the center of the Galaxy. 
    
 We assume that star cluster positions and velocities follow the same distribution as the stars in the host galaxy. Thus, their initial conditions are drawn accordingly to the galaxy distribution function of the energy, which in our case, corresponds to the aforementioned Dehnen model.
    This choice is not unique and it is intrinsically linked to the star cluster formation process and its dependence on the galaxy's structure and evolution. Still, it represents a good balance between the reliability of the initial conditions  and the computational cost of the simulation. 
A different distribution of the cluster's initial conditions would likely affect their overall evolution. For example, a more concentrated distribution of star clusters would favor a more significant amount of mass that is delivered into the Galactic center and likely a larger nuclear cluster mass. In contrast, a looser distribution would favor the tidal disruption of the clusters and the formation of a less massive and dense nuclear cluster \citep{manuel15}. 
    
    \section{Method}
  We studied the evolution throughout the simulation of five representative clusters of different masses and concentrations.
For the mass, we selected de clusters using the number of particles since the particle mass is constant ($M=92 M_\odot$).  For the concentration, we used the concentration parameter ($c$) defined as $c=\log (r_t/r_c)$ where $r_t$ is the tidal radius, that is, the value of the radius at which the density profile reaches zero and $r_c$ the core radius which is the radius where the density has dropped to half the central value \citep{king62}. 

   The final sample consists of $GC_3$ as the least massive, $GC_{31}$ as the intermediate-mass, and $GC_{23}$ as the most massive clusters.
    For the different concentrations, the clusters selected are $GC_3$, $GC_{26}$, and $GC_4$ for the least, intermediate, and most concentrated, respectively. We note that 
    $GC_3$ was used in both analyses as the least massive and concentrated cluster. 
    The five GCs selected are well distributed around the center (Fig.~\ref{fig:M_clblue}), which avoids any dependence on the initial position of the cluster that can influence the results. 
 Table~\ref{tab:M_tablesel} lists the five clusters selected along with the number of particles ($N$), total mass ($M_{GCs}$), Lagrange radius ($r_L$), Tidal radius ($r_t$), Core radius ($r_c$), and concentration parameter ($c$).

 \begin{table}
%\caption{References to the catalogs used for this work with their numbers of RRab stars and radial coverage for each catalog in degrees and parsecs.}  
\caption{Properties of the Globular clusters in the simulation according to mass ($GC_3$, $GC_{31}$, and $GC_{23}$) and concentration ($GC_3$, $GC_{26}$, and $GC_4$). $GC_3$ was used for both comparisons, Column 1: GC name. Column 2: Number of particles ($N$). Column 3: Total mass ($M_{GCs}$) in $10^6 \times M_\odot$. Column 4: Lagrange radius ($r_L$) in pc, Column 5: Tidal radius ($r_t$) in pc.  Column 6: Core radius ($r_C$) in pc. Column 7: Concentration parameter ($c$)}  
\label{tab:M_tablesel}    
\centering                  
\begin{tabular}{c c  c c c c c}      
\hline\hline          
     GC name & N & $M_{GC}$ & $r_L$ & $r_t$ & $r_c$ & $c$ \\
     & & {\tiny ($10^6 M_\odot$)} & {\tiny (pc)} & {\tiny (pc)} & {\tiny (pc)} & \\
\hline  
     $GC_{3}$ & 1477 & 0.13 & 2.14 & 11.73 & 1.08 & 2.38 \\ 
     $GC_{4}$ & 20700 & 1.90 & 1.47 & 12.66 & 0.56 & 3.11 \\
     $GC_{23}$ & 21383 & 1.96 & 3.20 & 27.16 & 0.86 & 3.44 \\
     $GC_{26}$ & 11117 & 1.02 & 2.30 & 18.92 & 0.85 & 3.09\\
     $GC_{31}$ & 11134 & 1.02 & 1.97 & 16.15 & 0.60 & 3.28\\
\hline 
\end{tabular}
\end{table}

    Figure~\ref{fig:M_spatial1} (mass comparison of $GC_3$, $GC_{31}$ and $GC_{23}$) and Fig.~\ref{fig:M_spatial2} (concentration comparison of $GC_3$, $GC_{26}$, and $GC_4$) present the spatial distribution of the clusters in the first snapshot, volumetric density profile, and Lagrange radius evolution.
    The differences in mass and concentration of the clusters selected for the comparison are clear from the spatial distribution and the density profile (left and middle panels of Fig.~\ref{fig:M_spatial1} and Fig.~\ref{fig:M_spatial2}).
    Although the $GC_{23}$ (lower panel of Fig.~\ref{fig:M_spatial1}) is the most massive cluster of the simulation, the cluster $GC_4$ (lower panel of Fig.~\ref{fig:M_spatial2}) is more concentrated, reaching a value of $\sim 10^{10} $ particles per deg$^3$.
    
     In Fig.~\ref{fig:M_evol1} (mass comparison of $GC_3$, $GC_{31}$ and $GC_{23}$) and Fig.~\ref{fig:M_evol2} (concentration comparison of $GC_3$, $GC_{26}$ and $GC_4$) a graphical representation of the spatial distribution of three representative snapshots at $t = 0$ Myr, $t = 8.3$ Myr, and $t = 41.5$ Myr are shown. We show three representative snapshots, but the complete evolution was used to analyze the evolution in morphology properly.
     The full animations of the GCs evolution are available in the online version of this paper.

 These simulations clearly show that effect of the supermassive BH is truly devastating. All GCs are destroyed (most of them very rapidly) and their debris mixed. These remains end up forming an extended structure around the supermassive BH that is akin to a nuclear star cluster. During the process, we observe the emergence of complex features in the individual GCs. When observed from different viewing angles, these features appear as threads, multiple blobs, shells, sausages, and so on. We also note that thin long, and coherent tidal tails, which are very common in disrupting halo GCs, are not present in these simulations.
    
    %Indirect observations of these disruption processes can be searched using hypervelocity stars that can be ejected by interaction with the supermassive BH as proposed by \cite{Hills88}. These hypervelocity stars are rare and were discovered in the Galactic halo \citep{Brown05, Brown15, Marchetti18}, and in the Magellanic Clouds \citep{Przybilla08, Lennon17}, but have also recently been observed in the inner Galactic bulge \citep{alonso19, evans22}. Furthermore, the disruption of clusters can produce multiple hypervelocity stars expelled contemporaneously from the Galactic center \citep{giacomo16}. Our simulations show multiple stars that escape the Galactic center region but very few hypervelocity objects.
      
    \section{Results}
\begin{figure*}[b]
    \begin{center}
    \includegraphics[width=0.8\linewidth]{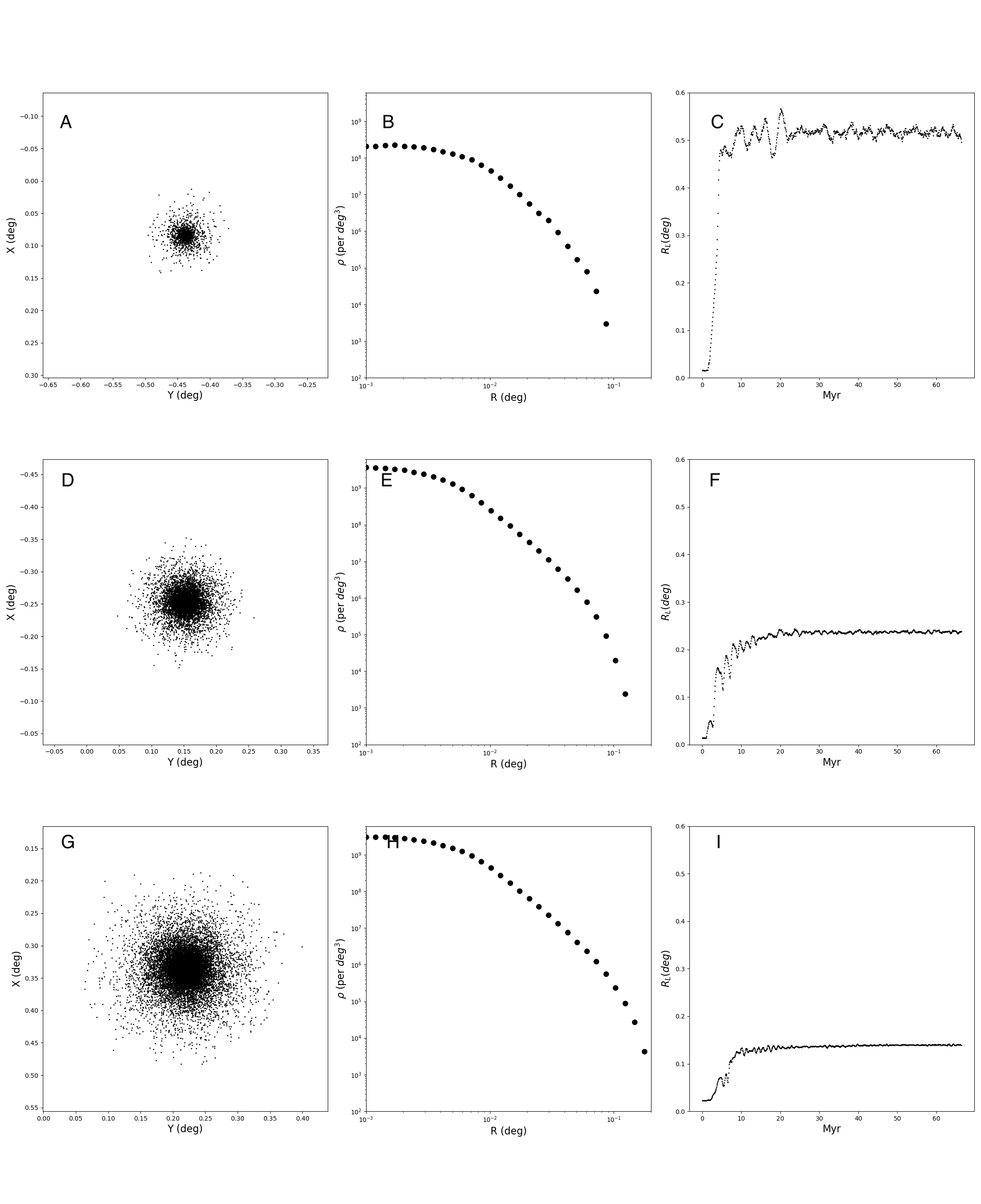}\\
    \caption[Spatial distribution, Volumetric density profile and Lagrange radius evolution for the three clusters selected by mass]
    {Globular clusters with different masses. 
Left:\ Spatial distribution of the particles belonging to the clusters in the first snapshot of the simulation.  Middle:\ Volumetric density profiles
for the clusters in the first snapshots of the simulation. Right:\ Implemented conversion from parsecs to degrees, assuming the Galactocentric
distance $R_0 = 8.2$ kpc \citep{grav17}. Right:\ Evolution of the $50\%$ Lagrange radius of the clusters during the simulation.
Top, middle, and lower panels show the least massive cluster ($GC_3$), the intermediate mass cluster ($GC_{31}$), and the most massive cluster ($GC_{23}$), respectively.
    \label{fig:M_spatial1}}
    \end{center}
    \end{figure*}

\begin{figure*}[b]
    \begin{center}
    \includegraphics[width=0.8\linewidth]{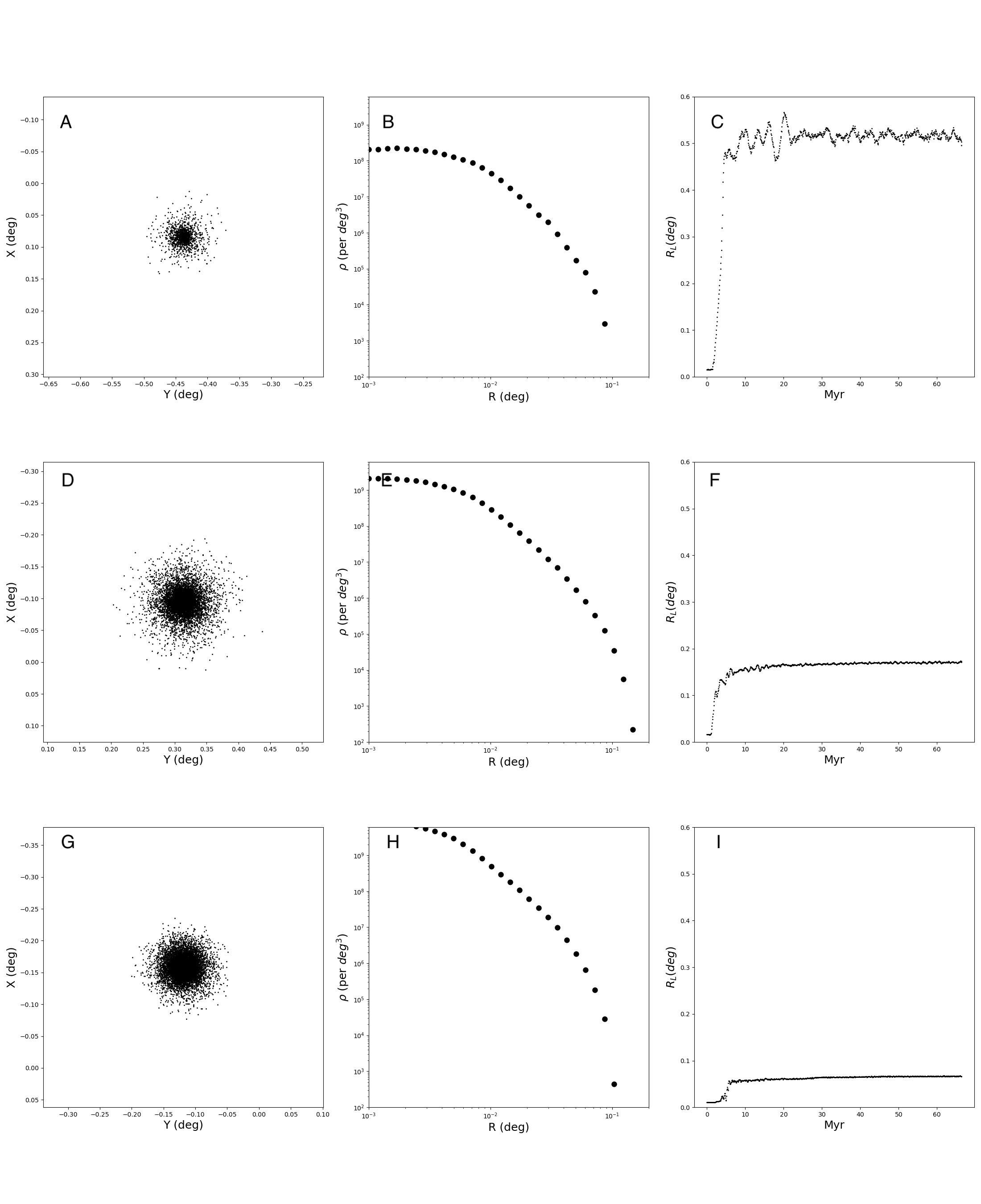}\\
    \caption[Spatial distribution, Volumetric density profile and Lagrange radius evolution for the three clusters selected by concentration]
    {
       Globular clusters with different concentrations. Left:\ Spatial distribution of the particles belonging to the clusters in the first snapshot of the simulation. Middle:\ Volumetric density profiles
for the clusters in the first snapshots of the simulation. Right:\ Implemented conversion from parsecs to degrees, assuming the Galactocentric
distance $R_0 = 8.2$ kpc. \citep{grav17}. Right:\ Evolution of the $50\%$ Lagrange radius of the clusters during the simulation.
Top, middle, and lower panels show the least massive cluster ($GC_3$), the intermediate mass cluster ($GC_{26}$), and the most massive cluster ($GC_{4}$), respectively.
    \label{fig:M_spatial2}}
    \end{center}
    \end{figure*}

\begin{figure*}[b]
    \begin{center}
    \includegraphics[width=0.8\linewidth]{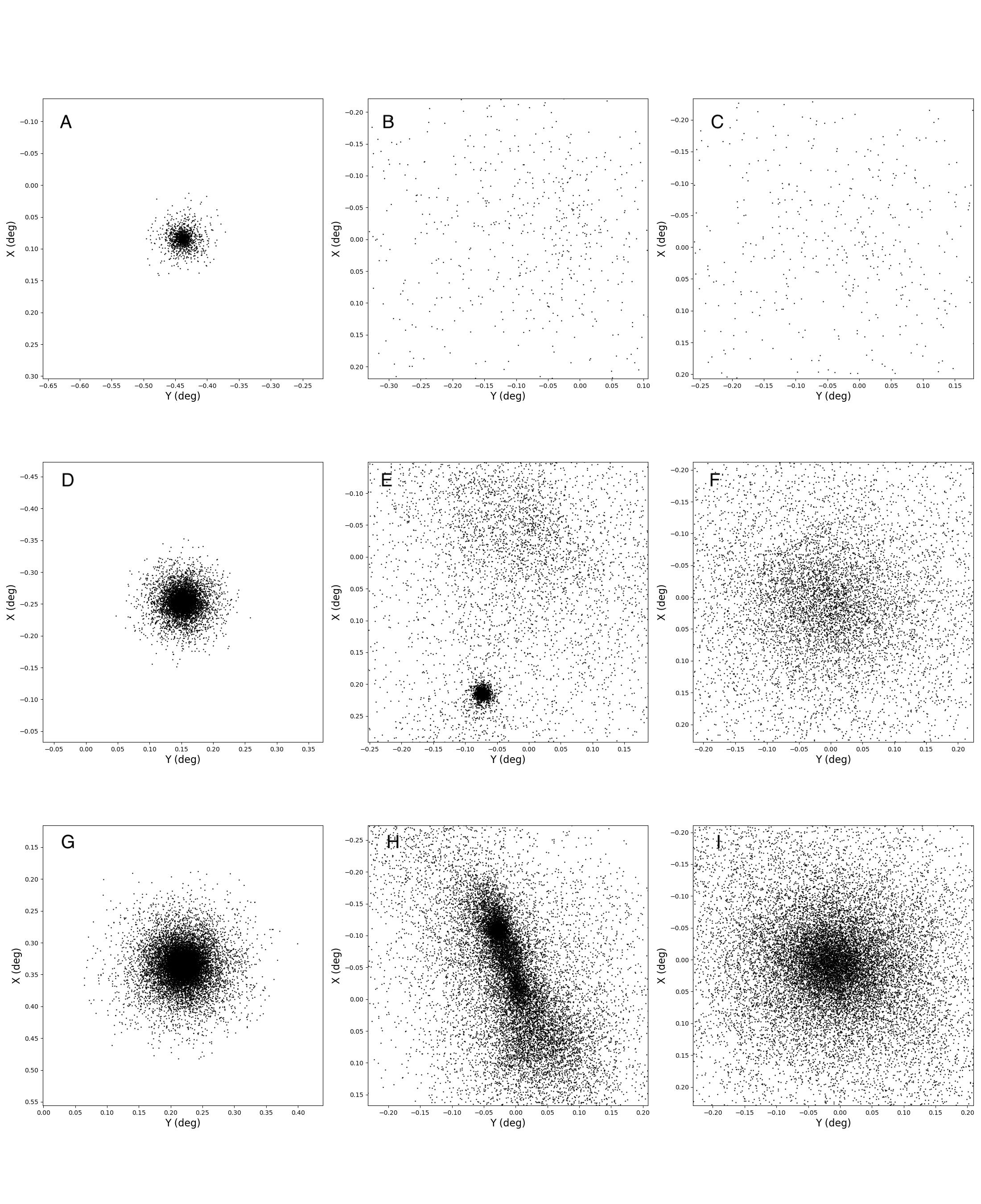}\\
    \caption[Snapshots of the three clusters selected by mass during the simulation]
    {
    Spatial distribution of the least massive cluster $GC_3$ during the simulation, shown at the top. Middle: Intermediate
mass cluster $GC_{31}$. Bottom: Most massive cluster $GC_{23}$.
Left plots show the\ first snapshot of the simulation ($t = 0$ Myr), while the middle and right plots correspond to $t = 8.3$ Myr and $t = 41.5$ Myr,
respectively. Complete animations for the simulations are available online.
    \label{fig:M_evol1}}
    \end{center}
    \end{figure*}

\begin{figure*}[b]
    \begin{center}
    \includegraphics[width=0.8\linewidth]{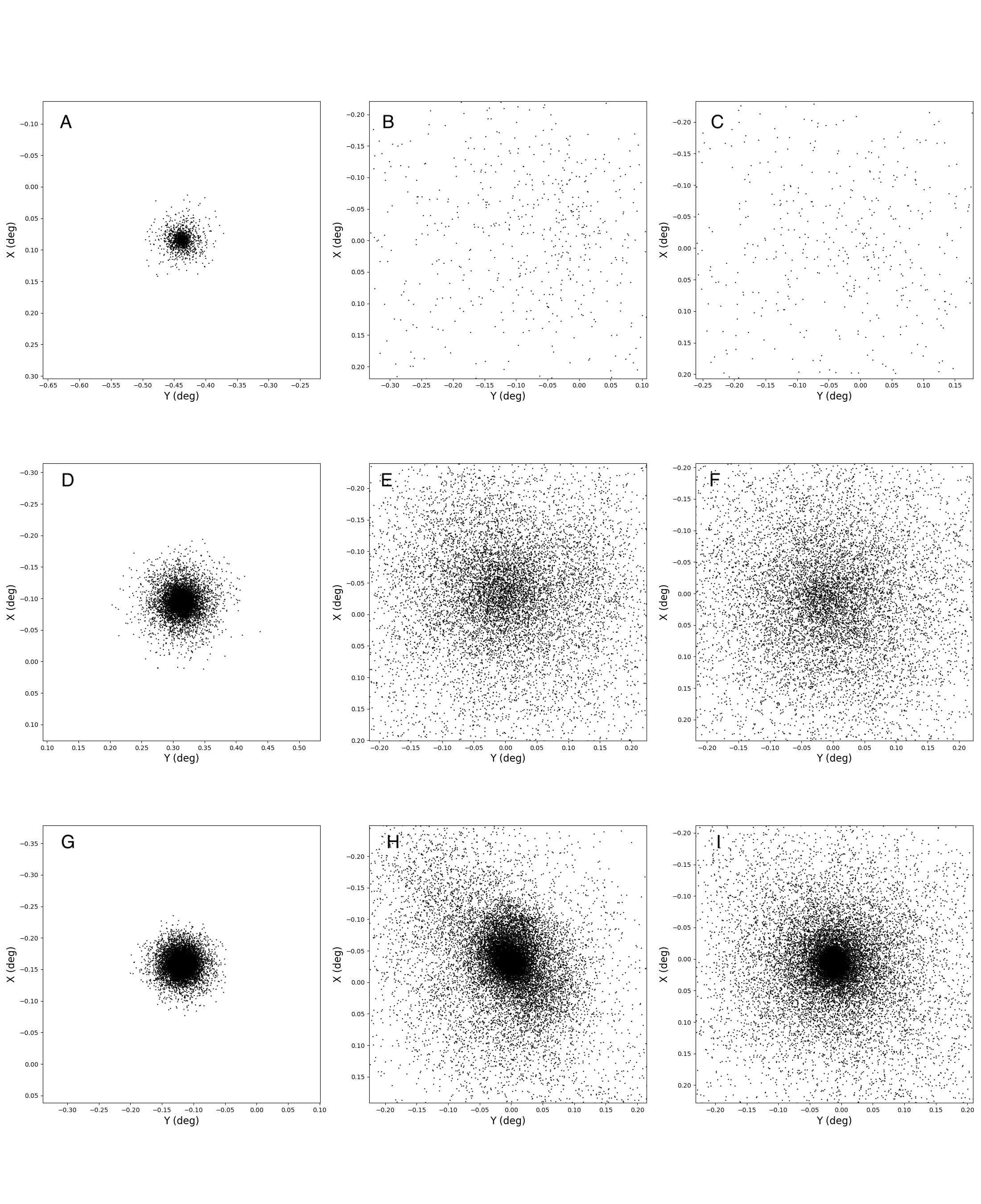}\\
    \caption[Snapshots of the three clusters selected by concentration during the simulation]
    {
    Spatial distribution of the least concentrated cluster $GC_3$ during the simulation, shown at the top. Middle:\ Intermediate concentration cluster $GC_{26}$. Bottom:    Most concentrated cluster $GC_4$. 
Left plots show the first snapshot of the simulation ($t = 0$ Myr), while the middle and right plots correspond to $t = 8.3$ Myr and $t = 41.5$ Myr, respectively. Complete animations for the simulations are available online.
    \label{fig:M_evol2}}
    \end{center}
    \end{figure*}

    The evolution of the different clusters can be analyzed from different points of view, as discussed in the following.
In terms of morphology and survival, we used two indicators to evaluate the evolution of the different clusters, the $50\%$ Lagrange radius (half-mass radius), as shown in the right panels of Figs.~\ref{fig:M_spatial1} and \ref{fig:M_spatial2}, and the graphical representation of the spatial distribution of the clusters during the complete simulation. The latter is displayed as three representative snapshots  in Figs.~\ref{fig:M_evol1} and \ref{fig:M_evol2}, whereas the full animations of the GCs evolution shown in these figures are available in the online version of this paper to illustrate the morphological changes.
    
    When analyzing the GCs of different masses (Figs.~\ref{fig:M_spatial1} and~\ref{fig:M_evol1}), the results show that the least massive cluster ($GC_{3}$) is completely destroyed and falls into the potential well very quickly. 
    Panel B of Fig.~\ref{fig:M_evol1} shows that the cluster is already completely disrupted at $t = 8.3$ Myr.
    This is also evident in the Lagrange radius evolution in panel C of Fig.~\ref{fig:M_spatial1}, where the cluster experiences prominent variations in the initial snapshots, reaching a value of $R = 0.511$ deg ($72.9$ pc) at 8 Myr.
    The intermediate-mass cluster ($GC_{31}$) keeps its high stellar concentration core during the first part of the simulation and then no longer has a defined shape.
    In panel E of Fig.~\ref{fig:M_evol1}, the core is still clearly defined at $t = 8.3$ Myr, but in panel F at $t = 41.5$ Myr, we can see that it is destroyed.
    The Lagrange radius of this cluster stabilizes at $\sim 19 Myr,$ without presenting significant variations in size, reaching a value of $R = 0.234$ deg ($33.3$ pc). 
    The most massive cluster ($GC_{23}$) instead undergoes deformations and elongations of the most central and concentrated part (see H panel of Fig.~\ref{fig:M_evol1}) and then it is destroyed, leaving an appreciable concentration of stars in the center.
    The Lagrange radius evolution shows a bump at $\sim 5$ Myr following its first passage near the Galactic center, where the potential is maximized (I panel of Fig.~\ref{fig:M_spatial1}).
    This means that the destructive effect of the supermassive BH is more critical than that of cluster-cluster collisions. 
    Then this massive GC takes $25.7$ Myr to reach stability at a value of $R = 0.136$ deg ($19.4$ pc). 
    
    On the other hand, the clusters of different concentrations exhibit different behavior  (Figs.~\ref{fig:M_spatial2} and~\ref{fig:M_evol2}). 
    The least concentrated cluster corresponds to the least massive already analyzed ($GC_{3}$).
    As mentioned, this cluster is disrupted quickly, spreading its stars into the field. 
    The intermediate concentration one ($GC_{26}$) does not keep the core for a long time. 
    According to panels D, E, and F of Fig.~\ref{fig:M_evol2}, at $t = 8.3$ Myr, the cluster no longer has a defined shape. 
    The Lagrange radius begins to stabilize at $17$ Myr but reaches stability at $36$ Myr, with a value of $R = 0.165$ deg ($23.5$ pc) (see F panel of Fig.~\ref{fig:M_spatial2}). 
    In the lower panels (G, H, and I ) of Fig.~\ref{fig:M_evol2}, the most concentrated cluster ($GC_{4}$) exhibits various changing shapes in its outskirts while the central core survives.
    It begins to stabilize at $30$ Myr and reaches stability at $42$ Myr with a Lagrange radius of only $R = 0.065$ deg ($9.2$ pc) (I panel of Fig.~\ref{fig:M_spatial2}).
    Therefore, the least concentrated clusters are destroyed faster than the more concentrated ones at a given mass.
    The clusters' morphological evolution and destruction during the simulation strongly depend on their mass and concentration.

    At the end of the simulation, a concentrated nuclear star cluster is formed, containing the bulk of the total GC mass. %(XX of the initial mass-- using four pc?). 
     Its evolution is discussed in detail by Navarro et al. in preparation.
 We observe that for low mass and low concentration clusters (panel C of Figs.~\ref{fig:M_spatial1} and~\ref{fig:M_spatial2}, respectively), the Lagrange radius stabilizes faster, reaching a higher value, meaning that the least massive and concentrated clusters disrupt fast and spread the particles around without yielding an evident concentration of stars around the Galactic center. 
    The high mass and concentrated clusters (panel C of Figs.~\ref{fig:M_spatial1} and~\ref{fig:M_spatial2}, respectively) reach smaller values for the Lagrange radius, meaning that the stars belonging to these clusters remain confined around the Galactic center after the simulation. This implies that the massive and concentrated GCs contribute to a higher percentage of stars in the Galactic center than the least massive GCs. Therefore, the population of stars concentrated in the Galactic center comes primarily from massive clusters. 
    
    It can also be observed that the massive clusters suffer deformations in their shapes (lower panels G, H, and I of Fig.~\ref{fig:M_evol1}). 
    The deformation can be general or limited to the outer particles depending on the concentration of the cluster.
    We see that the morphology is directly affected by the passages of the clusters near the supermassive BH.
    Interestingly, a close passage by the central BH can split the cluster into two well-separated structures that are not necessarily symmetric as we find, for example, in GC tidal tails.
    This is shown as bumps in the Lagrange radius evolution at the moment when these passages occur.
    The clusters may exhibit different morphologies during this process.
    
   Additionally, we studied the evolution of another sample of six clusters from the simulation, along with the parameters mentioned before.   
    The test in terms of mass was also carried out for the clusters $GC_{30}$, $GC_{25}$, and $GC_{14}$ as the second most massive, intermediate-mass, and least massive clusters.
    In terms of concentration, we also analyzed the clusters $GC_{10}$, $GC_{18}$, and $GC_{11}$ as the second most concentrated, intermediate, and least concentrated clusters.
        The behavior is similar to the clusters selected here ($GC_3$, $GC_4$, $GC_{23}$, $GC_{26}$, and $GC_{31}$) in terms of the morphology evolution and final distribution of particles depending on mass and concentration. %which means that the analysis is robust and does not depend on the initial position of the clusters.
    
    One of the motivations of our simulations has been to guide a future search for GCs hidden in the Galactic center region with the Roman Space Telescope. Still, this work implies that these GCs should be long gone within $\sim 300$ pc.
We thus chose to examine the specific example of VVV-GC002, the closest GCs to the Galactic center at $R_G = 400$ pc \citep{Minniti21}. 
Even though this distance range is just outside our simulations that reach $300$ pc, we can predict that this GC should not have survived in that environment.
Therefore, it either consists of the remains of a much more massive structure or it has formed farther away and is now observed close to the perigalactic of its orbit -- or both assumptions may be true. 
The GCs of our simulation are massive and some are comparable in mass to VVV-GC002. However, a direct comparison with the orbital evolution of this cluster is not yet warranted  because the orbit of VVV-GC002 is still unknown, as this GC lacks radial velocities  \citep{Minniti21}.

%Unfortunately, the VVV-GC002 orbital evolution cannot be computed because this GC has no RV measurements.

    \section{Conclusions}
We use N-body simulations based on the MEGaN project, consisting of the evolution of 41 Globular clusters falling into the potential well of the Milky Way nucleus as one of its main formation mechanisms. We analyzed the morphology evolution and final fate of five Globular clusters as a representative sample to evaluate the evolution dependence on mass and concentration. 
    As indicators, we use the clusters' morphology in all the simulation snapshots projected in the three axes, along with the spatial distribution, density profile, and the evolution of the $50\%$ Lagrange radius (half mass radius).
We study the behavior of $GC_3$, $GC_{31}$, and $GC_{23}$ as the least, intermediate, and most massive clusters, respectively. 
    Using the concentration parameter ($c$) defined as $c=\log (r_t/r_c)$ as an indicator, we selected $GC_3$ as the least concentrated cluster, $GC_{26}$ for the intermediate concentrated one, and $GC_4$ for the most concentrated cluster. 
    
    From the Lagrange radius evolution, we see that while the looser globular clusters are destroyed very quickly (<10 Myr), some can survive a few passages close to the supermassive BH before completely disappearing. Instead, the most massive clusters survive for more extended periods before being wholly disrupted. During their evolution, they show variations in morphology depending on their concentration and the distance to the central supermassive BH. The least concentrated clusters are more likely to show essential variations in their shape. Instead, the more concentrated ones remain in a quasi-spherical shape during the complete simulation, showing superficial deformations in their shapes, particularly from the outer particles.

    At the end of the simulation, the particles coming from the most massive and concentrated clusters are confined in the inner region; instead, the least massive and concentrated clusters get disrupted early and spread particles at distances longer to the center. This means that the stars from massive and concentrated primordial globular clusters have contributed a higher percentage to the nuclear star cluster than those from lower mass clusters. 
    
   % Those survivors are preferentially massive, concentrated clusters in low eccentricity orbits.  
    %We find that the least massive clusters get totally destroyed in the first passage close to the nuclear BH, lasting least than one orbit. The most massive ones are able to remain longer, lasting several orbits, and their cores remain as coherent structures orbiting around the supermassive BH for a long time.
    The Galactic center region is difficult to explore in detail from the ground due to extreme crowding and high extinction. However, it could be mapped most efficiently at high resolution in the near-infrared with the wide-field camera of the Roman Space Telescope (WFIRST, \citealt{Green12, Spergel15}). We could then search for dissolving clusters and their remaining cores and carry out a complete census of the stellar populations composing the nuclear star clusters to carry out a comparison with the  simulations presented here.


\begin{thebibliography}{}
\bibitem[Aarseth et al. (1974)]{Aarseth74} Aarseth, S. J., Hénon, M., Wielen, R., 1974, A\&A, 37, 183
\bibitem[Aarseth (1999)]{Aarseth99} Aarseth, S. J. 1999, PASP, 111, 1333
\bibitem[Arca-Sedda \& Capuzzo-Dolcetta (2014)]{Arca14} Arca-Sedda, M., \& Capuzzo-Dolcetta, R. 2014, ApJ, 785, 51
\bibitem[Arca-Sedda et al.(2015)]{manuel15} Arca-Sedda, M., \& Capuzzo-Dolcetta, R., Antonini, F. \& Seth, A., 2015, ApJ, 806, 220
\bibitem[Arca-Sedda \& Capuzzo-Dolcetta (2017)]{manuel17} Arca-Sedda, M. \& Capuzzo-Dolcetta, R., 2017, MNRAS, 471
\bibitem[Arca-Sedda \& Capuzzo-Dolcetta (2018)]{manuel18} Arca-Sedda, M. \& Capuzzo-Dolcetta, R., 2018, MNRAS, 483
\bibitem[Baugh et al. (1996)]{Baugh96} Baugh, C. M., Cole, S., \& Frenk, C. S. 1996, MNRAS, 283, 1361
\bibitem[Baumgardt et al. (2002)]{Baumgardt02} Baumgardt, H., Hut, P., \& Heggie, D. C. 2002, MNRAS, 336, 1069
\bibitem[Baumgardt \& Makino(2003)]{Baumgardt03} Baumgardt, H. \& Makino, J. 2003, MNRAS, 340, 227
\bibitem[Brown et al.(2005)]{Brown05} Brown, W. R., Geller, M. J., Kenyon, S. J., \& Kurtz, M. J. 2005, ApJL, 622, L33
\bibitem[Brown et al.(2015)]{Brown15} Brown, W. R. 2015, ARA\&A, 53, 15
\bibitem[Carlberg(2017)]{Carlberg17} Carlberg, R. G., 2017, ApJ, 838, 39 
\bibitem[Carlberg (2018)]{Carlberg18} Carlberg, R. G. 2018, ApJ, 861, 69C
\bibitem[Capuzzo-Dolcetta (1993)]{rcd93}  Capuzzo-Dolcetta, R., 1993, ApJ, 415, 616
\bibitem[Chandrasekhar (1942)]{Chandrasekhar42} Chandrasekhar, S. 1942, ”Stellar Dynamics”, Univ. of Chicago Press
\bibitem[Chernoff \& Weinberg (1990)]{Chernoff90} Chernoff D. F., \& Weinberg M. D., 1990, ApJ, 351, 121
\bibitem[Dehnen (1993)]{Dehnen93} Dehnen, W., 1993, MNRAS, 265, 250
\bibitem[Eggen, Lynden-Bell \& Sandage(1962)]{ELS} Eggen, O. J., Lynden-Bell, D., Sandage, A. R., 1962, ApJ, 136, 748 
\bibitem[Evans et al. (2022)]{evans22} Evans, F., Marchetti, T. \& Rossi, E., 2022, MNRAS, 517, 3
\bibitem[Fall \& Rees (1977)]{Fall77} Fall, S. M. \& Rees, M. J. 1977, MNRAS, 181, 37
\bibitem[Fall \& Rees (1985)]{Fall85} Fall, S. M. \& Rees, M. J. 1985 Astrophys. J., 298, 18
\bibitem[Fragione \& Capuzzo-Dolcetta (2016)]{giacomo16} Fragione, G., \& Capuzzo-Dolcetta, R. 2016, 458, 2596F
\bibitem[Garro et al. (2022a)]{elisa22a} Garro, E., Minniti, D., Alessi, B., et al.,  A\&A, 659, 155, 18
\bibitem[Garro et al. (2022b)]{elisa22b} Garro, E., Minniti, D., Gomez, M., et al.,  A\&A, 662, 95, 18
\bibitem[Gieles et al. (2006)]{Gieles06} Gieles M., Portegies Zwart S. F., Baumgardt H., et al., 2006, MNRAS, 371, 793
\bibitem[Gnedin \& Ostrike (1997)]{Gnedin97} Gnedin, O. Y., \& Ostriker, J. P. 1997, ApJ. 474, 223
\bibitem[Gnedin et al. (2014)]{Gnedin14} Gnedin, O. Y., Ostriker, J. P., \& Tremaine, S. 2014, ApJ, 875, 71
\bibitem[Green et al.(2012)]{Green12}  Green, J., Schechter, P., Baltay, C., et al. 2012, arXiv:1208.4012
\bibitem[Habibi et al. (2013)]{Habibi13} Habibi, M., Stolte, A., Brandner, W., Hussmann, B, \& Motohara, K. 2013, A\&A, 556, A26
\bibitem[Habibi et al. (2014)]{Habibi14} Habibi, M., Stolte, A., \& Harfst, S. 2014, A\&A, 566, A6
%\bibitem[Harris, (1996)]{Harris96} Harris, W. E., 1996, AJ, 112, 1487
\bibitem[Harris, (1996)]{harris10} Harris, W. E., 2010, arXiv 1012.3224
\bibitem[Heggie (1979)]{Heggie79} Heggie, D. C. 1979, MNRAS, 76, 525
\bibitem[Heggie (2014)]{Heggie14} Heggie, D. C. 2014, MNRAS, 445, 3435
\bibitem[H\'enon (1961)]{Henon61} H\'enon, M. 1961, Annales d'Astrophysique, 24, 369
\bibitem[Hills(1988)]{Hills88} Hills, J. G. 1988, Nature, 331, 687
\bibitem[Hosek et al. (2015)]{Hosek15} Hosek M.W., Lu J.R., Anderson J., Ghez A.M., Morris M.R., \& Clarkson W.I., 2015, ApJ, 813, 27
\bibitem[Khoperskov et al.(2018)]{Khoperskov18} Khoperskov, S., Mastrobuono-Battisti, A., Di Matteo, P., \& Haywood, M. 2018, A\&A, 620, A154
\bibitem[King (1962)]{king62} King, I. 1962, AJ, 67, 471
\bibitem[Larson (1970)]{Larson70} Larson, R. B. 1970, MNRAS, 147, 323
\bibitem[Lennon et al.(2017)]{Lennon17} Lennon, D., van der Marel, R. P., Ramos Lerate, M., et al. 2017, A\&A, 603, A75
\bibitem[Libralato et al. (2020)]{Libralato20} Libralato, M., Fardal, M., Lennon, D., et al. 2020, MNRAS, 497, 4
\bibitem[Libralato et al. (2022)]{Libralato22} Libralato, M., Bellini, A., Vesperini, E., et al. 2022, ApJ, 934, 2
\bibitem[Luna et al.(2019)]{alonso19} Luna, A., Minniti, D., \& Alonso-Garcia, J. 2019, ApJL, 887, L39
\bibitem[Marchetti et al.(2018)]{Marchetti18} Marchetti, Contigiani, O., Rossi, E. M., et al. 2018, MNRAS, 476, 4697
\bibitem[Minniti(1995)]{d95} Minniti, D. 1995, AJ, 109, 1663
%\bibitem[Minniti et al.(2016)]{dante16} Minniti D., Contreras Ramos, R., Zoccali, M., et al. 2016, ApJL , 830, 14
\bibitem[Minniti et al.(2010)]{vvvminniti10} Minniti D., Lucas, P. W., Emerson, J. et al. 2010, New Astron., 15, 433
\bibitem[Minniti et al.(2017)]{dante17} Minniti D.,  Geisler, D., Alonso-García, J., et al. 2017, ApJL , 849, 24
\bibitem[Minniti et al. (2021)]{Minniti21} Minniti, D., Fernández-Trincado, J. G, Smith, L. C., Lucas, P. W., Gomez, M., \& Pullen, J. B. 2021, A\&A, 648A, 86
\bibitem[Moni Bidin et al. (2011)]{Moni11} Moni Bidin, C., Mauro, F., Geisler, D., 2011, A\&A, 535, A33
\bibitem[Navarro et al.(2021)]{navarro21} Navarro, M. G., Minniti, D., Capuzzo-Dolcetta, R., et al. 2021, A\&A, 646, 45
\bibitem[Neistein et al. (2006)]{Neistein06} Neistein, E., van den Bosch, F. C., \& Dekel, A. 2006, MNRAS, 372, 933
\bibitem[Nitadori \& Makino(2008)]{Nitadori08} Nitadori, K. \& Makino, J., 2008, NewA, 13, 498 
\bibitem[Przybilla et al.(2008)]{Przybilla08} Przybilla, N., Nieva, M. F., Heber, U., et al. 2008, A\&A, 480, L37
\bibitem[Capuzzo-Dolcetta et al.(2013)]{rcd13} R. Capuzzo-Dolcetta, M. Spera, D. Punzo, 2013., Journ. of Comp. Phys., v. 236, p. 580-593
\bibitem[Rui et al. (2019)]{Rui19} Rui, N. Z., Hosek, M. W. Jr., Lu, J. R., et al. 2019, ApJ, 877, 37
\bibitem[Spergel et al.(2015)]{Spergel15} Spergel, D., Gehrels, N., Baltay, C., et al. 2015, arXiv:1503.03757
\bibitem[Skrutskie et al.(2006)]{Skrutskie06} Skrutskie, M.F., Cutri, R.M., Stiening, R., et al. 2006, AJ, 131, 1163
\bibitem[The GRAVITY Collaboration (2019)]{grav17} The GRAVITY Collaboration, 2019, A\&A, 625, L10 
\bibitem[Tremaine et al. (1975)]{tre75} Tremaine, S. D., Ostriker, J. P., \& Spitzer, L. 1975, ApJ, 196, 407
\bibitem[Tremaine \& Weinberg (1984)]{Tremaine84} Tremaine, S., \& Weinberg, M. D. 1984, MNRAS, 209, 729
\bibitem[Vesperini \& Heggie (1997)]{Vesperini97} Vesperini, E., \& Heggie, D. C. 1997, MNRAS, 289, 898
\bibitem[White \& Rees (1978)]{wr78} White, S. D. M. \& Rees, M. J., 1978, MNRAS, 183, 431
\end{thebibliography}
\end{document}